\documentclass[12pt]{amsart}

\usepackage{cite} % generates citation ranges
\usepackage{color} % standard color package
\usepackage{graphicx} % standard graphics package
\usepackage{latexsym,amssymb,amsmath,amsfonts} % extended symbol packages
\usepackage{mathrsfs} % mathscripts \mathscr
\usepackage{fixltx2e} % improves figure floating in Latex2e kernel

\allowdisplaybreaks[4]
\newtheorem{mylemma}{Lemma}[section]
\newtheorem{mydefinition}{Definition}[section]
\newtheorem{mytheorem}{Theorem}[section]
\newtheorem{mycorollary}{Corollary}[section]
\newtheorem{myexample}{Example}[section]

\definecolor{Light}{gray}{0.85}

\def\abs#1{\left\vert #1 \right\vert}
\def\allpolyx0degn{\mbox{$P_n$}}

\def\bfem#1{{\bf \em #1}} % boldface italics

\def\bull{\rule{0.08in}{0.08in}} % square filled bullet

\def\C{{\mathbb C}} % complex numbers (AMS symbol)
\newcommand{\comment}[1]{} % allows one to comment out a block of text

\def\eqref#1{(\ref{#1})} % parentheses around referenced equation numbers
\def\mbf#1{\hbox{\mathversion{bold}$#1$}} % math boldface
\def\mbfss#1{\hbox{\scriptsize\mathversion{bold}$#1$}} % math boldface
\def\nat{{\mathbb N}} % natural numbers (AMS symbol)

\def\openbull{\framebox[0.08in][c]{$\;$}} % square unfilled bullet

\def\re{{\mathbb R}} % real numbers (AMS symbol)

\def\shuffle{{\scriptscriptstyle \;\sqcup \hspace*{-0.05cm}\sqcup\;}}

\def\begals{\[\begin{aligned}}
\def\endals{\end{aligned}\]}
\def\begce{\begin{center}}
\def\endce{\end{center}}
\def\begar{\begin{array}}
\def\endar{\end{array}}
\def\begeq{\begin{equation}}
\def\endeq{\end{equation}}
\def\begdi{\begin{displaymath}}
\def\enddi{\end{displaymath}}
\def\begdis{\begin{eqnarray*}}
\def\enddis{\end{eqnarray*}}
\def\begeqa{\begin{eqnarray}}
\def\endeqa{\end{eqnarray}}
\def\begdes{\begin{description}}
\def\enddes{\end{description}}
\def\begit{\begin{itemize}}
\def\endit{\end{itemize}}
\def\begen{\begin{enumerate}}
\def\enden{\end{enumerate}}
\def\beglar{\left[\begin{array}}
\def\endrar{\end{array}\right]}
\def\begle{\begin{mylemma}}
\def\endle{\end{mylemma}}
\def\begde{\begin{mydefinition}}
\def\endde{\end{mydefinition}}
\def\begth{\begin{mytheorem}}
\def\endth{\end{mytheorem}}
\def\begco{\begin{mycorollary}}
\def\endco{\end{mycorollary}}
\def\begprop{\begin{myproposition}}
\def\endprop{\end{myproposition}}
\def\begex{\begin{myexample} \rm}
\def\endex{\hfill\openbull \end{myexample} \vspace*{0.15in}}
\def\begexer{\begin{myexercise}}
\def\endexer{\end{myexercise}}

\def\begres{\noindent{\bf Remarks}:\begin{enumerate}}
\def\endres{\end{enumerate} \par}
\def\begpr{\noindent{\em Proof:}$\;\;$}
\def\endpr{\hfill\bull \vspace*{0.15in}}
\def\begtab{\begin{tabular}}
\def\endtab{\end{tabular}}
\def\rref#1{(\ref{#1})}
\def\shuff#1#2{\mathbin{
      \hbox{\vbox{\hbox{\vrule \hskip#2 \vrule height#1 width 0pt}\hrule}\vbox{\hbox{\vrule \hskip#2 \vrule height#1 width 0pt\vrule }\hrule}}}}
\def\shuffl{{\mathchoice{\shuff{5pt}{3.5pt}}{\shuff{5pt}{3.5pt}}{\shuff{3pt}{2.6pt}}{\shuff{3pt}{2.6pt}}}}
\def\shuffle{{\, \shuffl \,}}
\def\Dseries#1{{\mathscr D}_{#1}\langle\langle X\rangle\rangle}
\def\DseriesXc#1{{\mathscr D}_{#1}\langle\langle X_c\rangle\rangle}
\def\DseriesXd#1{{\mathscr D}_{#1}\langle\langle X_d\rangle\rangle}\def\opercomp{\bullet} % operator composition

\begin{document}

\title[Parameter Dependent Chen--Fliess Series]
{Parameter Dependent Chen--Fliess Series and \\ Their Nonrecursive Interconnections}

\author{W. Steven Gray}
\address{Department of Electrical and Computer Engineering, Old Dominion University, Norfolk, Virginia 23529, USA}
\email{sgray@odu.edu}
\urladdr{http://www.ece.odu.edu/~sgray}
\date{}

\author{Natalie Pham}
\address{Department of Electrical and Computer Engineering, Old Dominion University, Norfolk, Virginia 23529, USA}
\email{npham003@odu.edu}
\date{}

\begin{abstract}
A class of parameter dependent Chen--Fliess series is introduced where the series coefficients are taken from a noncommutative ring of multivariable differential operators.
Such series are shown in the linear case to represent formal solutions to Cauchy initial value problems for nonhomogeneous PDEs and thus are useful for characterizing the input-output maps of distributed control systems. It is also shown that this class of functional series is {\em almost} closed under the set of nonrecursive interconnections, that is, any finite combination of parallel and series interconnections without a closed-loop.
Some sufficient conditions are needed for the series interconnection.
Specific examples are given involving the transport equation and the wave equation.
\end{abstract}

\maketitle

\noindent
\tableofcontents


\vspace{0.1in}


\thispagestyle{empty}

\section{Introduction}

Chen--Fliess functional series, which are infinite series of scaled iterated integrals, constitute a particularly convenient class
of nonlinear input-output systems due to their representations
in terms of noncommutative formal power series and all the various algebraic structures available in this
setting to support computations \cite{Fliess_81,Fliess_83,Gray-etal_SCL14,Kawski-Sussmann_97}.
In addition, their convergence properties are well
understood \cite{Fliess_81,Gray-Wang_02 ,Thitsa-Gray_SIAM12,Venkatesh-Gray_23,Winter-Arboleda_19},
they have a well developed realization theory \cite{Fliess_83,Isidori_95,Jakubczyk_80,Jakubczyk_86,Jakubczyk_86a,Sussmann_77,Nijmeijer-vanderSchaft_90,Sussmann_90}, and their relationship to input-output differential equations
has been extensively characterized \cite{Sontag-Wang_89,Wang-Sontag_92,Wang-Sontag_92a,Wang-Sontag_95}. As a class, they are closed under all the usual types of system interconnections encountered in
applications \cite{Ferfera_80,Fliess_81,Foissy_15,Gray-etal_SCL14,Gray-Li_SIAM05}.

In this paper,
a class of {\em parameter dependent} Chen--Fliess series is introduced where the series coefficients are taken from a noncommutative ring of multivariable differential operators
rather than from a field as done in the classical case (usually either $\re$ or $\C$).
Such series have the potential to represent formal solutions to Cauchy initial value problems for nonlinear nonhomogeneous PDEs.
The concept is being employed implicitly in most inductive schemes for solving such problems (see, for example, \cite{Parker-Sochacki_00}).
In this initial development of the
concept, the focus will be primarily on the linear case where the required algebraic machinery for doing computations is already available.
Of course, such PDE's are often used to characterize the input-output maps of a distributed control systems. So its applications are immediate.
It should also be noted that the generalization present here is distinct from that given in \cite{Duffaut_Espinosa-etal_18}, where the coefficients
are from a field but instead the iterated integrals are matrix-valued and thus not commutative.

It is next shown that this class of functional series is {\em almost} closed under the set of nonrecursive interconnections, that is, any finite combination of parallel and series interconnections without a closed-loop.
Some sufficient conditions are needed for the series interconnection. This represents the first significant departure from the classical case.
Specific examples are then given involving the transport equation and the wave equation. Finally, feedback connections are {\em not} considered here as they
will likely require a major effort to determine under what conditions such interconnections are even well defined.
This is a direct consequence of the fact that the algebraic nature of the series interconnection is more complex in this new setting, and feedback interconnections are ultimately comprised of an infinite sequence of series interconnections.
Thus, that topic is deferred to future work.

The paper is organized as follows. In the next section, the general definition of a parameter dependent Chen--Fliess series is presented. It is then specialized to
the linear case in order to provide some sufficient conditions for convergence. In Section~\ref{sec:Nonrecursive-System-Interconnections},
the parallel and series connections are then considered. The conclusions of the paper are summarized in the final section.

\section{Parameter Dependent Chen--Fliess Series}
\label{sec:PD-CFS}

The general case is first presented. Then the single-input, single-output (SISO) linear case is further developed.

\subsection{General Case}

Fix an alphabet $X=\{x_0,x_1,\ldots,x_m\}$. Let $X^\ast$ denote the set of all words over $X$ including the
empty word $\emptyset$.
For each $i\in\{1,\ldots,m\}$,
let $u_i$ be a real-valued function defined on $\Theta\times [t_0,t_0+T]$, where $\Theta\subseteq\re^{\rm d}$ with
${\rm d}\in\nat$, $t_0\in\re$, and $T\in\re^+$ fixed.
It will often be assumed that $t_0=0$. In addition, $u$ is presumed to be a smooth function of the
parameters $\theta=(\theta_1,\ldots,\theta_{\rm d})\in \Theta$, and
for every fixed $\theta$ assume that $u_i(\theta,\cdot)\in L_1[t_0,t_0+T]$.
A {\em Chen series} in this setting is defined to be the formal power series
\begdi
P(\theta,t)=\sum_{\eta\in X^\ast} E_\eta[u](\theta,t)\eta,
\enddi
where $E_\eta$ is defined inductively by
\begdi
E_{x_i\bar{\eta}}[u](\theta,t) = \int_{t_0}^tu_{i}(\theta,\tau)E_{\bar{\eta}}[u](\theta,\tau)\,d\tau
\enddi
with $E_\emptyset:=1$,
$x_i\in X$, $\bar{\eta}\in X^{\ast}$, and $u_0:=1$. Perfectly
analogous to the standard case \cite{Chen_54,Chen_57,Chen_58,Chen_77}, it follows that
\begdi
\frac{\partial P}{\partial t}=\left(\sum_{i=0}^m u_ix_i\right) P,\;\;P(\theta,t_0)=\mbf{1},
\enddi
where $\mbf{1}:=1\emptyset$.

Let $\alpha=(\alpha_1,\ldots,\alpha_d)$ be a multiindex with $\alpha_i\in\nat_0$. For each $\alpha$ there corresponds
a differential operator
\begdi
D_\alpha=\frac{\partial^{\alpha_1}}{\partial^{\alpha_1}\theta_1}\cdots \frac{\partial^{\alpha_{\rm d}}}{\partial^{\alpha_{\rm d}}\theta_{\rm d}}.
\enddi
Define $\abs{\alpha}=\sum_{i=1}^{\rm d}\alpha_i$.
Consider the ring $R$ of smooth $\re^\ell$ vector-valued functions on $\Theta$ with multiplication defined componentwise.
Then the noncommutative ring of multivariate polynomial differential operators in ${\rm d}$ variables over $R$ is
denoted by
${\mathscr D}_{\rm d}:=R[ D_1,\ldots,D_{\rm d}]$, where $D_i:=\partial/\partial\theta_i$.
If $R$ is replaced with $\re[\theta_1,\ldots,\theta_d]$ in the definition, then  ${\mathscr D}_{\rm d}$ becomes the $d$-th Weyl algebra \cite{Dixmier_68}.
For any word $\eta\in X^\ast$,
let $(c,\eta)\in{\mathscr D}_{\rm d}$ so that
\begdi
(c_i,\eta)=\sum_{\abs{\alpha}\leq n_\eta} a_\alpha D_\alpha,\;\;a_{\alpha,i}\in C^{\infty}(\Theta),\;\;i\in\{1,\ldots,\ell\},\;\;n_\eta\in \nat_0.
\enddi
Let $\Dseries{\rm d}$ denote the linear space of formal power series over $X$ with coefficients in ${\mathscr D}_{\rm d}$.
The following definition presents the central object of study in this paper.

\begde
An $m$-input, $\ell$-output \bfem{parameter dependent Chen--Fliess series} with generating series
$c\in \Dseries{\rm d}$ is defined by the scalar product
\begin{align*}
y_i(\theta,t)&=F_{c_i}[u](\theta,t)=\sum_{\eta\in X^\ast} (c_i,\eta)(\theta)(P[u](\theta,t),\eta) \\
&=\sum_{\eta\in X^\ast} (c_i,\eta)(\theta)E_\eta[u](\theta,t),\;\;i=1,\ldots,\ell.
\end{align*}
\endde

Of course, the summation above is only formal unless convergence can be proved. This issue is addressed
next in the linear setting.
% zzz
%\begth
%\textcolor[rgb]{1.00,0.00,0.00}{General convergence theorems}
%\endth

\subsection{Single-Parameter SISO Linear Case}

The single-parameter single-input, single-output case corresponds to setting ${\rm d}=1$, $m=1$, and $\ell=1$ in the
construction above. In which case, $X=\{x_0,x_1\}$, $R=C^\infty(\Theta)$, and $D_1$ is equivalent to the commutative
Weyl algebra when $R$ is replaced with $\re[\theta]$.
To linearize the setup,
define the following two subsets of $X^\ast$:
\begdi
X_0^\ast=\{\eta=x_0^k:k\geq 0\},\;\;X_L=\{\eta\in X^\ast: \abs{\eta}_{x_1}=1\},
\enddi
where $\abs{\eta}_{x_1}$ denotes the number of times the letter $x_1$ appears in $\eta$.
The linearized Chen series is then
\begdi
P_L[u]:=P_L^0[u]+P_L^1[u],
\enddi
where
\begdi
P_L^0[u]=\sum_{\eta\in X_0^\ast} E_\eta[u]\eta,\;\; P_L^1[u]=\sum_{\eta\in X_L} E_\eta[u]\eta.
\enddi

\begle \label{le:2D-Chen-series}
The series $P_L^0$, $P_L^1$, and $P_L$ satisfy the following linear equations, respectively:
\begin{align*}
\frac{\partial}{\partial t}P_L^0[u]&=x_0P_L^0[u] \\
\frac{\partial}{\partial t}P_L^1[u]&=x_0P_L^1[u]+u_1x_1P_L^0[u] \\
\frac{\partial}{\partial t}P_L[u]&=x_0P_L[u]+u_1x_1P_L^0[u].
\end{align*}
\endle

\begpr
All the identities follow directly from the following calculation:
\begin{align*}
\frac{\partial}{\partial t}P_L[u]&=\frac{\partial}{\partial t}P_L^0[u]+\frac{\partial}{\partial t}P_L^1[u] \\
&=\frac{\partial}{\partial t}\sum_{\eta\in X_0^\ast}\eta E_{\eta}[u]+\frac{\partial}{\partial t}\sum_{\eta\in X_L}\eta E_{\eta}[u] \\
&=\sum_{\eta\in X_0^\ast}x_0\eta E_{\eta}[u]+\sum_{\eta\in X_L}x_0\eta E_{\eta}[u]+u_1\sum_{\eta\in X_0^\ast}x_1\eta E_{\eta}[u] \\
&=x_0P_L[u]+u_1x_1P_L^0[u].
\end{align*}
\endpr

\begde
A SISO \bfem{parameter dependent linear Chen--Fliess series} with generating series
$c\in \Dseries{1}$ is defined by the scalar product
\begeq \label{eq:linearized-CFS}
y(\theta,t)=F_{c,L}[u](\theta,t)=\sum_{\eta\in X_0^\ast\cup X_L} (c,\eta)(\theta)(P_L[u](\theta,t),\eta).
\endeq
\endde

The following lemma will be useful for providing sufficient conditions under which the series \rref{eq:linearized-CFS} converges.

\begle \label{le:E-bound-linear-words}
Fix $a,b,T\in\re$ with $a<b$.
Let $u$ be smooth in $\theta\in\Theta=[a,b]$ and assume that for every $k\in\nat_0$
\begdi
    \left\|\frac{\partial^ku}{\partial \theta^k}\right\|_1
    =\int_{a}^{b}\int_{0}^{T}\left|\frac{\partial^k}{\partial \theta^k}u(\theta,\tau)\right|\,d\tau\,d\theta
\enddi
is finite.
Then given any $i,j\in\nat_0$, it follows that
\begdi
    \left|E_{x_0^ix_1x_0^j}\left[\frac{\partial^ku}{\partial \theta^k}\right](\theta,t)\right|\leq
    \frac{i^ij^jT^{i+j}}{i!j!(i+j)^{i+j}}\left\|\frac{\partial^ku}{\partial \theta^k}\right\|_1
\enddi
for all $(\theta,t)\in[a,b]\times[0,T]$.
\endle

\begpr
From H\"{o}lder's inequality
\begin{align*}
   \left|E_{x_0^ix_1x_0^j}\left[\frac{\partial^ku}{\partial \theta^k}\right](\theta,t)\right|&=\left|\int_{0}^{t}\frac{(t-\tau)^i}{i!}\frac{\partial^k}{\partial \theta^k}u(\theta,\tau)\frac{\tau^j}{j!}\,d\tau\right| \\
    &\leq\int_{a}^{b}\int_{0}^{T}\left|\frac{(T-\tau)^i}{i!}\frac{\partial^k}{\partial \theta^k}u(\theta,\tau)\frac{\tau^j}{j!}\right|\,d\tau\,d\theta \\
    &\leq \left\|\frac{(T-\tau)^i}{i!}\frac{\tau^j}{j!}\right\|_\infty \left\|\frac{\partial^k u}{\partial \theta^k}\right\|_1 \\
    & \leq \frac{i^ij^jT^{i+j}}{i!j!(i+j)^{i+j}}  \left\|\frac{\partial^k u}{\partial \theta^k}\right\|_1.
\end{align*}
\endpr

\begco \label{co:E-bound-linear-words}
There exists a constant $K_E>0$ such that
\begdi
    \left|E_{x_0^ix_1x_0^j}\left[\frac{\partial^ku}{\partial \theta^k}\right](\theta,t)\right|\leq
    K_E\frac{T^{i+j}}{(i+j)!}\left\|\frac{\partial^ku}{\partial \theta^k}\right\|_1
\enddi
for all $i,j\in\nat_0$ and $(\theta,t)\in[a,b]\times[0,T]$.
\endco

\begpr
The claim is a direct result of Lemma~\ref{le:E-bound-linear-words} and
Stirling's formula $i^i/i!\sim {\rm e}^i/\sqrt{2\pi i}$, $i\gg 1$.
\endpr

The next theorem provides sufficient conditions under which the zero state
portion of \rref{eq:linearized-CFS} converges when
\begdi
(c,x_0^ix_1x_0^j)=\alpha_{i+j}\frac{\partial^{i+j}}{\partial\theta^{i+j}}.
\enddi
In short, if certain growth rates on $\|\alpha_k\|_\infty$ and $\|\partial^k u/\partial \theta^k\|_1$ apply
as a function of $k$, then the linear Chen--Fliess series converges provided that $[a,b]$ and $[0,T]$ are sufficiently small in length.
That is, these conditions ensures convergence in a local sense.

\begth
Suppose that $\Theta=[a,b]$ is a finite interval so that each $\alpha_{i+j}\in L_{\infty}[a,b]$.
Assume  there exist real numbers $K_\alpha,M>0$ satisfying
\begdi
    \|\alpha_{k}\|_\infty\leq K_\alpha M^{k}k!,\;\;\forall k\in\nat_0.
\enddi
Let $u$ be smooth in $\theta$ with $\partial^k u/\partial \theta^k\in L_1([a,b]\times[0,T])$ for every $k\in \nat_0$. Assume there exist
real numbers $K_u,R>0$ so that
\begdi
    \left\|\frac{\partial^{k}u}{\partial\theta^{k}}\right\|_1\leq K_uR^{k},\;\;\forall k\in \nat_0.
\enddi
Then the parameter dependent linear Chen--Fliess series
\begdi
    F[u](\theta,t)=\sum_{i,j=0}^{\infty}\alpha_{i+j}(\theta)\frac{\partial^{i+j}}{\partial\theta^{i+j}}E_{x_0^ix_1x_0^j}[u](\theta,t)
\enddi
converges absolutely and uniformly on $[a,b]\times[0,T]$ when $MRT<1$.
\endth

\begpr
It follows from the stated assumptions and Corollary~\ref{co:E-bound-linear-words} that
\begin{align*}
    |F[u](\theta,t)|&\leq \sum_{i,j=0}^\infty\|\alpha_{i+j}\|_\infty\left|\frac{\partial^{i+j}}{\partial\theta^{i+j}}E_{x_0^ix_1x_0^j}[u](\theta,t)\right|\\
    &\leq\sum_{i,j=0}^{\infty} K_\alpha M^{i+j}(i+j)!\; K_E\frac{T^{i+j}}{(i+j)!}\left\|\frac{\partial^{i+j}u}{\partial \theta^{i+j}}\right\|_1 \\
    &=K_\alpha K_E K_u\sum_{i,j=0}^{\infty}(MRT)^{i+j}\\
    &=K_\alpha K_E K_u\left(\frac{1}{1-MRT}\right)^2
\end{align*}
provided that $MRT<1$.
\endpr

The following theorem introduces a more restrictive growth condition on the series coefficients so that the condition $MRT<1$ above is no longer required.
That is, $[a,b]$ and $[0,T]$ need only be finite. This is viewed as a type of global convergence.

\begth
Suppose that $\Theta=[a,b]$ is a finite interval so that each $\alpha_{i+j}\in L_{\infty}[a,b]$.
Assume  there exist real numbers $K_\alpha,M>0$ satisfying
\begdi
    \|\alpha_{k}\|_\infty\leq K_\alpha M^{k}(k!)^s,\;\;\forall k\in\nat_0,
\enddi
where $0\leq s <1$.
Let $u$ be smooth in $\theta$ with $\partial^k u/\partial \theta^k\in L_1([a,b]\times[0,T])$ for every $k\in \nat_0$. Assume there exist
real numbers $K_u,R>0$ so that
\begdi
    \left\|\frac{\partial^{k}}{\partial\theta^{k}}u\right\|_1\leq K_uR^{k},\;\;\forall k\in \nat_0.
\enddi
Then the parameter dependent linear Chen--Fliess series
\begdi
    F[u](\theta,t)=\sum_{i,j=0}^{\infty}\alpha_{i+j}(\theta)\frac{\partial^{i+j}}{\partial\theta^{i+j}}E_{x_0^ix_1x_0^j}[u](\theta,t)
\enddi
converges absolutely and uniformly on $[a,b]\times[0,T]$.
\endth

\begpr
Similar to the proof of the previous theorem, observe
\begin{align*}
    |F[u](\theta,t)|&\leq \sum_{i,j=0}^\infty\|\alpha_{i+j}\|_\infty\left|\frac{\partial^{i+j}}{\partial\theta^{i+j}}E_{x_0^ix_1x_0^j}[u](\theta,t)\right|\\
    &\leq\sum_{i,j=0}^{\infty} K_\alpha M^{i+j}((i+j)!)^s\; K_E\frac{T^{i+j}}{(i+j)!}\left\|\frac{\partial^{i+j}u}{\partial \theta^{i+j}}\right\|_1 \\
    &=K_\alpha K_E K_u\sum_{i,j=0}^{\infty}(MRT)^{i+j}\frac{((i+j)!)^s}{(i+j)!}\\
    &=K_\alpha K_E K_u\sum_{k=0}^{\infty}(k+1)(MRT)^k\frac{(k!)^s}{k!}.
\end{align*}
The final step above follows from the standard double summation formula $\sum_{i,j=0}^\infty b_{i,j}=\sum_{k=0}^\infty \sum_{l=0}^k b_{l,k-l}$, where now $b_{l,k-l}$
is invariant with respect to $l$.
Define the sequence $a_k=(k+1)(MRT)^k(k!)^s/(k!)$, $k\in \nat_0$ and observe that
\begdi
    \lim_{k\rightarrow\infty}\left|\frac{a_{k+1}}{a_k}\right|=
    MRT\lim_{k\rightarrow\infty}\frac{k+2}{k+1}\cdot\frac{1}{(k+1)^{1-s}}
    =0
\enddi
since $0\leq s<1$.
Thus, from the ratio test, the series converges absolutely and uniformly on $[a,b]\times[0,T]$.
\endpr

\begex
Consider a single-parameter SISO system $u\mapsto y$ defined in terms of the transport equation
\begeq \label{eq:transport-equation}
\frac{\partial y}{\partial t}+V\frac{\partial y}{\partial \theta}=u,
\endeq
where $(\theta,t)\in\re\times\re^+$ and $V\in C^{\infty}(\re)$.
Assume the smooth initial condition
$y(\theta,0)=y_0(\theta)$.
The assertion is that the formal solution to this linear PDE
can be written in the form $y=F_{c,L}[u]$ for
some suitable generating series $c\in \Dseries{1}$ .
To determine this series, first observe from Lemma~\ref{le:2D-Chen-series}
that
\begin{align*}
\frac{\partial y}{\partial t}&=\left(c,\frac{\partial}{\partial t}P_L[u]\right) \\
&=(c,x_0P_L[u])+u (c,x_1P_L^0[u]),
\end{align*}
or equivalently,
\begdi
\frac{\partial y}{\partial t}-(x_0^{-1}(c),P_L[u])=u (x_1^{-1}(c),P_L^0[u]).
\enddi
(For brevity $u_1$ is written here as $u$.)
Comparing this identity against the transport equation yields the system of equations:
\begin{align}
(x_1^{-1}(c),P_L^0[u])&=1 \label{eq:c-system-eq1}\\
-(x_0^{-1}(c),P_L[u])&=V\frac{\partial y}{\partial \theta}. \label{eq:c-system-eq2}
\end{align}
Since $y=(c,P_L[u])$, it also follows that
\begdi
y(\theta,0)=(c,P_L[u](\theta,0))=(c,\emptyset)(\theta)=y_0(\theta).
\enddi
Equation \rref{eq:c-system-eq1} has the explicit form
\begdi
\sum_{k=0}^\infty (c,x_1x_0^k)\frac{t^k}{k!}=1,\;\;\forall t\geq 0.
\enddi
Thus,
\begdi
c=y_0+\sum_{k=1}^\infty (c,x_0^k)x_0^k+x_1+\sum_{k=1}^\infty (c,x_0^kx_1)x_0^kx_1.
\enddi
It follows from \rref{eq:c-system-eq2} that
\begdi
-(x_0^{-1}(c),P_L^0[u])-(x_0^{-1}(c),P_L^1[u])=V\frac{\partial y}{\partial \theta},
\enddi
where
\begdi
V\frac{\partial y}{\partial \theta}=
V\left(\frac{\partial c}{\partial \theta},P_L^0[u]\right)+V\left(\frac{\partial c}{\partial \theta},P_L^1[u]\right)+V\left(c,\frac{\partial}{\partial \theta}P_L^1[u]\right).
\enddi
Thus, one must conclude that
\begin{align*}
(c,x_0^{k+1})&=-V\frac{\partial}{\partial \theta}(c,x_0^k) \\
(c,x_0^{k+1}x_1)&=-V\frac{\partial}{\partial \theta}(c,x_0^kx_1)-V(c,x_0^kx_1)\frac{\partial}{\partial \theta}.
\end{align*}
Therefore,
\begdi
c=\sum_{k=0}^\infty \left(-V\frac{\partial}{\partial \theta}\right)^ky_0\,x_0^k+
\sum_{k=0}^\infty \left(-V\frac{\partial}{\partial \theta}\right)^k x_0^kx_1,
\enddi
which yields the Chen--Fliess series solution to the transport equation
\begeq \label{eq:CF-solution-transport-eq}
y=\sum_{k=0}^\infty \left(-V\frac{\partial}{\partial \theta}\right)^k y_0 E_{x_0^k}[u]+\sum_{k=0}^\infty \left(-V\frac{\partial}{\partial \theta}\right)^k E_{x_0^kx_1}[u].
\endeq
This series solution of the transport equation was computed in \cite{Pham-Gray_22} via state space methods. A somewhat simpler approach is
given in Section~\ref{sec:p-CFS-series-connection}.
If, for example, $V$ is constant and $u(\theta,t)=t\sin(\omega\theta)$, then it follows that
\begin{align*}
y(\theta,t)&=\sum_{k=0}^\infty \left.\frac{d^k}{d t^k}(y_0(\theta-Vt))\right|_{t=0}\frac{t^k}{k!}+\sum_{k=0}^\infty (-V)^k \frac{t^{k+2}}{(k+2)!}\frac{d^k}{d \theta^k}\sin(\omega\theta) \\
&=y_0(\theta-Vt)+\frac{\sin(\omega(Vt-\theta))+\sin(\omega \theta)-V\omega t\cos(\omega \theta)}{(V\omega)^2}.
\end{align*}
From Theorem 2.3, the series converges globally with growth constants
$K_\alpha=1$, $M=|V|$, $K_u=T^2(b-a)/2$, $R=|\omega|$, and $s=0$.
\endex

\section{Nonrecursive System Interconnections}
\label{sec:Nonrecursive-System-Interconnections}

In this section, the nonrecursive interconnections of parameter dependent Chen--Fliess series are characterized.
The  most frequently encountered of such interconnections in applications are the parallel and
series interconnections. The focus here will primarily be on the SISO case. The main two questions to be addressed are: Under what conditions is this
class of systems closed under a given interconnection topology, and, when appropriate, what is the  generating series of the composite system?
As will be seen, these are purely algebraic problems when the operators are viewed as formal series.
In addition, there are also important analytical problems that arise. For example, are these interconnections {\em well-posed} in the
sense that all the inputs and outputs in a given network are well defined functions when the operators are convergent in some sense?
Furthermore, is the composite system convergent given that its subsystems
are convergent? Is the sense of convergence preserved? For example, if the subsystems are all globally convergent, is the
composite system also globally convergent or perhaps only locally convergent?
In the classical case, these issues were addressed comprehensively in \cite{Thitsa-Gray_SIAM12,Venkatesh-Gray_23,Winter-Arboleda_19}.
Such problems tend to be quite technical and outside the scope of the present paper.

\subsection{Parallel Connections}

\begin{figure}[tb]
\begin{center}
\includegraphics[scale=0.4]{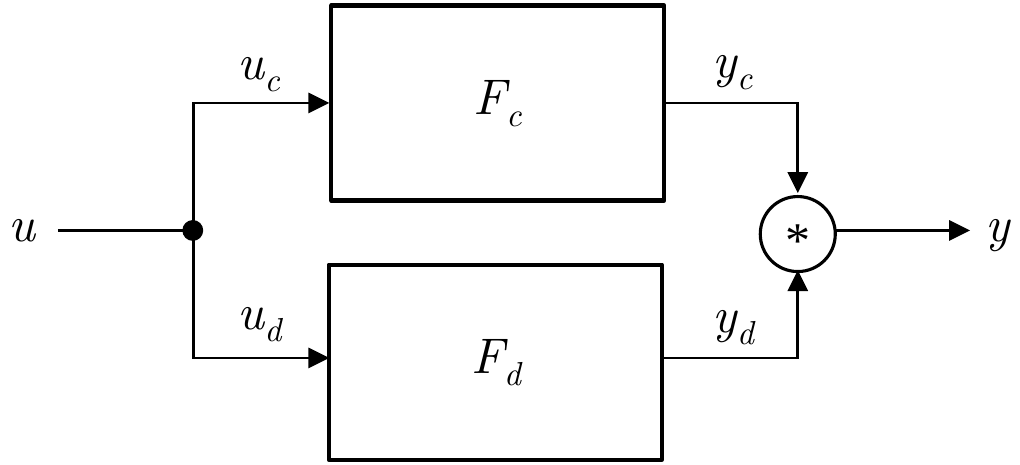}
\caption{Parallel interconnection of $F_c$ and $F_d$.}
\label{fig:parallel-connection}
\end{center}
\end{figure}

Consider two systems $F_c$ and $F_d$ interconnected in parallel as shown in Figure~\ref
{fig:parallel-connection}.
They have a common input $u$ and the outputs are combined by some binary operation $\ast$ to give a new output $y$.
The main theorem in this section provides two instances where the composite system can be
described by a parameter dependent Chen--Fliess series, namely, the
parallel sum and parallel product connections.
The proofs are analogous to the
classical case \cite{Fliess_81}. However, the shuffle product used for the parallel product connection
has to be handled carefully. The shuffle product
on words in $X^\ast$ is defined inductively by
\begdi
	(x_i\eta)\shuffle(x_j\xi)=x_i(\eta\shuffle(x_j\xi))+x_j((x_i\eta)\shuffle \xi),
\enddi
where $x_i,x_j\in X$, $\eta,\xi\in X^\ast$ and with $\eta\shuffle\emptyset=\emptyset\shuffle\eta=\eta$.
In the case where $c$ and $d$ are formal power series over $X^\ast$ with real coefficients, the definition is then
extended linearly so that
\begin{align*}
c\shuffle d&=\left(\sum_{\eta\in X^\ast} (c,\eta)\eta\right)\shuffle \left(\sum_{\xi\in X^\ast}(d,\xi)\xi\right) \\
&=\sum_{\eta,\xi\in X^\ast} (c,\eta)(d,\xi)\,\eta\shuffle\xi.
\end{align*}
If $c,d\in\Dseries{\rm d}$, however,
this last step is problematic as it no longer provides an algebra morphism describing the pointwise product in time $F_cF_d$
as illustrated by the following simple example.

\begex
Let $c=d=\partial/\partial \theta\, x_1$. Suppose
\begdi %\label{eq:bad-shuffle}
c\shuffle d=\frac{\partial}{\partial \theta}x_1\shuffle \frac{\partial}{\partial \theta}x_1=2\frac{\partial^2}{\partial \theta^2} x_1^2
\enddi
as suggested above.
In contrast, observe
\begin{align*}
F_c[u]F_d[u]&=\frac{\partial}{\partial \theta}E_{x_1}[u] \frac{\partial}{\partial \theta}E_{x_1}[u] \\
&=E_{x_1}\left[\frac{\partial}{\partial \theta}u\right]E_{x_1}\left[\frac{\partial}{\partial \theta}u\right] \\
&=F_{x_1\shuffle x_1}\left[\frac{\partial}{\partial \theta}u\right] \\
&=2F_{x_1^2}\left[\frac{\partial}{\partial \theta}u\right] \\
&\neq 2\frac{\partial^2}{\partial \theta^2}F_{x_1^2}[u].
\end{align*}
However, if $c=\partial/\partial \theta_1\, x_1$ and $d=\partial/\partial \theta_2\, x_2$,
then the usual linear extension of the shuffle product works as desired. Define $u=[u_1\;u_2]$
and $\theta=[\theta_1\;\theta_2]$
with $u_i$ only a function of $\theta_i$. Then
\begeq \label{eq:good-shuffle}
c\shuffle d=\frac{\partial}{\partial \theta_1}x_1\shuffle \frac{\partial}{\partial \theta_2}x_2
=\frac{\partial}{\partial \theta_1}\frac{\partial}{\partial \theta_2} x_1\shuffle x_2
=\frac{\partial}{\partial \theta_1}\frac{\partial}{\partial \theta_2}(x_1x_2+x_2x_1)
\endeq
and using integration by parts gives
\begin{align*}
F_c[u](\theta_1,t)F_d[u](\theta_2,t)&=\frac{\partial}{\partial \theta_1}E_{x_1}[u](\theta_1,t) \frac{\partial}{\partial \theta_2}E_{x_2}[u](\theta_2,t)\\
&=\int_0^t \frac{\partial u_1}{\partial \theta_1}(\theta_1,\tau)\,d\tau \int_0^t \frac{\partial u_2}{\partial \theta_2}(\theta_2,\tau)\,d\tau \\
&=\int_0^t \frac{\partial u_1}{\partial \theta_1}(\theta_1,\tau_1)\int_0^{\tau_1} \frac{\partial u_2}{\partial \theta_2}(\theta_2,\tau_2)\,d\tau_2d\tau_1 \\
&\hspace*{0.4in}+\int_0^t \frac{\partial u_2}{\partial \theta_2}(\theta_2,\tau_1)\int_0^{\tau_1} \frac{\partial u_1}{\partial \theta_1}(\theta_1,\tau_2)\,d\tau_2d\tau_1 \\
&=\frac{\partial}{\partial \theta_1}\frac{\partial}{\partial \theta_2}F_{x_1x_2+x_2x_1}[u](\theta,t) \\
&=\frac{\partial}{\partial \theta_1}\frac{\partial}{\partial \theta_2}F_{x_1\shuffle x_2}[u](\theta,t) \\
&=F_{c\shuffle d}[u](\theta,t),
\end{align*}
which is the result expected from \rref{eq:good-shuffle}. Thus, a certain {\em disjointness} is needed between the supports of  $c$ and $d$ in order for the
shuffle product to produce the desired algebra morphism. Fortunately, this feature is available for the parallel product connection, but as will be
seen later, it is not always available for series connections.
\endex

The main theorem of this section is now presented followed by a physical example.

\begth \label{th:pd-CFS-parallel-generating-series}
Define the alphabets $X_c=\{x_0,x_c\}$, $X_d=\{x_0,x_d\}$, and $X=X_c\cup X_d$.
Let $c\in\DseriesXc{{\rm d}_c}$ and $d\in\DseriesXd{{\rm d}_d}$ and define the
associated systems $y_c=F_c[u_c]$ and $y_d=F_d[u_d]$.
The parallel sum interconnection
\begdi
y(\theta,t)=F_c[u](\theta_c,t)+F_d[u](\theta_d,t)
\enddi
with $\theta:=(\theta_c,\theta_d)\in \Theta_c\times\Theta_d$ has the representation $y=F_{c+d}[u]$
with $c+d\in\Dseries{{\rm d}_c+{\rm d}_d}$.
Likewise,
the parallel product interconnection
\begdi
y(\theta,t)=F_c[u](\theta_c,t)F_d[u](\theta_d,t)
\enddi
has the representation $y=F_{c\shuffle d}[u]$ with $c\shuffle d\in\Dseries{{\rm d}_c+{\rm d}_d}$.
\endth

% zzz
%\begpr
%\textcolor[rgb]{1.00,0.00,0.00}{Need to rigorously prove the parallel product connection.}
%\endpr
%
%The corresponding convergence theorem is presented next for the case of linear operators.
%
%\begth
%zzz
%\endth
%
%\begpr
%zzz
%\endpr

\begex \label{ex:parallel-transport-systems}
Consider two single-parameter SISO transport systems
\begin{align*}
y_c(\theta_c,t)=F_c[u](\theta_c,t)&=\sum_{k=0}^\infty \left(-V_c\frac{\partial}{\partial \theta_c}\right)^k E_{x_0^kx_c}[u](\theta_c,t) \\
y_d(\theta_d,t)=F_d[u](\theta_d,t)&=\sum_{l=0}^\infty \left(-V_d\frac{\partial}{\partial \theta_d}\right)^l E_{x_0^lx_d}[u](\theta_d,t).
\end{align*}
The parallel sum connection is characterized by the linear generating series
\begdi
c+d=\sum_{k=0}^\infty
\left(-V_c\frac{\partial}{\partial \theta_c}\right)^k x_0^kx_c +
\left(-V_d\frac{\partial}{\partial \theta_d}\right)^k {x_0^kx_d}.
\enddi
It constitutes a transport system in the sense that
\begin{align*}
\frac{\partial y}{\partial t}&=\frac{\partial y_c}{\partial t}+\frac{\partial y_d}{\partial t}\\
&=\left(-V_c\frac{\partial y_c}{\partial \theta_c}+u\right)+\left(-V_d\frac{\partial y_d}{\partial \theta_d}+u\right) \\
&=-[V_c\;\;V_d]\frac{\partial y}{\partial\theta}+2u.
\end{align*}
The parallel product connection has the generating series
\begdi
c\shuffle d=\sum_{k,l=0}^\infty
\left(-V_c\frac{\partial}{\partial \theta_c}\right)^k
\left(-V_d\frac{\partial}{\partial \theta_d}\right)^l x_0^kx_c \shuffle {x_0^lx_d},
\enddi
which is no longer linear, and thus, does not represent a linear transport system.
\endex

\subsection{Series Connections}
\label{sec:p-CFS-series-connection}

\begin{figure}[tb]
\begin{center}
\includegraphics[scale=0.4]{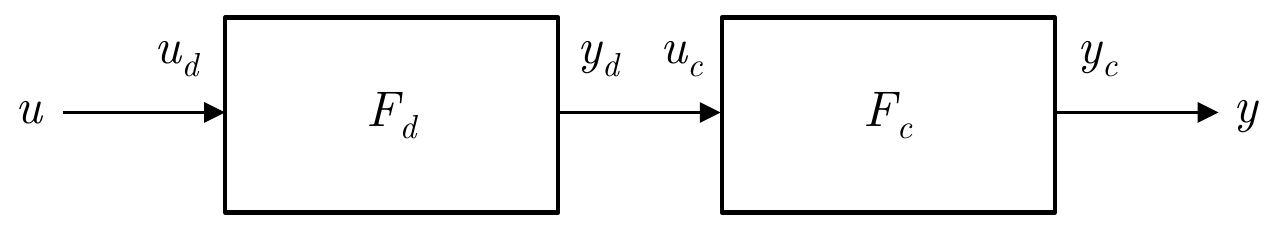}
\caption{Series interconnection of $F_c$ and $F_d$.}
\label{fig:series-connection}
\end{center}
\end{figure}

Consider next the series interconnection of $F_c$ and $F_d$ as shown in Figure~\ref{fig:series-connection}.
The interconnection is only well defined
when the output of $F_d$ is an {\em admissible} input to $F_c$.
For example, in a purely algebraic context, $u_c(\theta_c,t)=y_d(\theta_d,t)$ would be well defined if $\theta_c$ is identified with $\theta_d$.
A few simple examples are presented first to illustrate the mechanics of computing the generating series
for the composite system. Then a sufficient condition is given under which the series interconnection has a generating series in the sense
defined in Section~\ref{sec:PD-CFS}.

\begex \label{ex:simple-linear-composition}
Consider the composition of two single-parameter linear systems with generating series $c(\theta_c)=\theta_c(\partial/\partial\theta_c)x_c$ and $d(\theta_d)=\theta_d^2(\partial/\partial\theta_d)x_d$ so that
\begin{align*}
y_c(\theta_c,t)&=F_c[u_c](\theta_c,t)=\theta_c\frac{\partial}{\partial\theta_c}\int_0^t u_c(\theta_c,\tau)\,d\tau \\
y_d(\theta_d,t)&=F_d[u_d](\theta_d,t)=\theta_d^2\frac{\partial}{\partial\theta_d}\int_0^t u_d(\theta_d,\tau)\,d\tau.
\end{align*}
Setting $\theta=\theta_c=\theta_d\in[a,b]$, $u_d(\theta,t)=u(\theta,t)$, $u_c(\theta,t)=y_d(\theta,t)$, and $y(\theta,t)=y_c(\theta,t)$ gives
\begin{align*}
y(\theta,t)&=\theta\frac{\partial}{\partial\theta}\int_0^t y_d(\theta,\tau) \,d\tau \\
&=\theta\frac{\partial}{\partial\theta}\theta^2\frac{\partial}{\partial\theta}
\int_0^t \int_0^{\tau_1} u(\theta,\tau_2)\,d\tau_2\,d\tau_1 \\
&=(c,x_c)(\theta)(d,x_d)(\theta)E_{x_0x_d}[u](\theta,t) \\
&=F_{c\circ d}[u](\theta,t),
\end{align*}
where $c\circ d:=(c,x_c)(d,x_d)\, x_0x_d\in\DseriesXd{{1}}$.
Note that linearity is preserved.
\endex

\begex
Now modify the previous example so that $F_c$ is no longer linear, namely,
$c=(c,x_c^2)x_c^2$ and $d=(d,x_d)x_d$. Then
\begin{align*}
y(\theta,t)&=(c,x_c^2)(\theta)E_{x_c^2}[y_d](\theta,t) \\
&=(c,x_c^2)(\theta)\int_0^t y_d(\theta,\tau)E_{x_c}[y_d](\theta,\tau)\,d\tau,
\end{align*}
where
\begin{align*}
E_{x_c}[y_d](\theta,\tau)&=E_{x_c}[F_d[u]](\theta,\tau) \\
&=\int_0^\tau F_d[u](\theta,s)\,ds \\
&=\sum_{\eta\in X_d}(d,\eta)(\theta)\int_0^\tau E_{\eta}[u](\theta,s)\,ds \\
&=\sum_{\eta\in X_d}(d,\eta)(\theta)E_{x_0\eta}[u](\theta,\tau) \\
&=F_{x_0d}[u](\theta,\tau).
\end{align*}
Therefore,
\begin{align*}
y(\theta,t)&=(c,x_c^2)(\theta)\int_0^t F_d[u](\theta,\tau)F_{x_0d}[u](\theta,\tau)\,d\tau \\
&=(c,x_c^2)(\theta)\int_0^t F_{d\shuffle x_0d}[u](\theta,\tau)\,d\tau \\
&= (c,x_c^2)(\theta)F_{x_0(d\shuffle x_0d)}[u](\theta,\tau).
\end{align*}
This would lead one to believe that $c\circ d=(c,x_c^2)\,x_0(d\shuffle x_0d)$, which actually is the correct
result in the standard (nonparametric) theory. But as discussed in the previous section, the shuffle product on $\Dseries{1}$ alone does
not provide the desired algebra morphism without some modification. For example, one could
expand the alphabet for $d$ as $X_d=\{x_0,x_d(1),x_d(2)\}$ and then reparameterize the shuffle product so that
$x_0(d\shuffle x_0d)$ becomes $x_0(d_1\shuffle x_0d_2)$ with $d_1$ having its support in $\{x_0,x_d(1)\}^\ast$, and $d_2$ has
its support contained in $\{x_0,x_d(2)\}^\ast$. In general, this would require $X_d$ to be an infinite alphabet unless
$c$ is {\em input-limited}. That is, there exists an integer $N_c\in\nat$ such that for all $\eta\in X_c^\ast$,
$\abs{\eta}_{x_c}\leq N_c$. Interestingly, this is a known sufficient condition in the standard theory for the
composition
product to preserve rationality \cite{Ferfera_80}.
\endex

Fortunately, the standard SISO definition of the composition product generalizes directly when
$F_c$ is linear since $c$ is input-limited with $N_c=1$.
First, fix $d\in\DseriesXd{{\rm d}}$ and define the linear operators
\begin{align*}
\psi_d(x_0)&:\DseriesXd{{\rm d}}\rightarrow \DseriesXd{{\rm d}}, e\mapsto x_0e \\
\psi_d(x_c)&:\DseriesXd{{\rm d}}\rightarrow \DseriesXd{{\rm d}}, e\mapsto x_0(d\shuffle e).
\end{align*}
Extend the definition inductively to words in $X_c^\ast$ by letting
\begdi
\psi_d(x_ix_j)=\psi_d(x_i)\opercomp \psi_d(x_j),
\enddi
where `$\opercomp$' above denotes operator composition, and $\psi_d(\emptyset)$ is the identity map.
Define next a composition product on $X_c^\ast\times \DseriesXd{{\rm d}}$ by setting
\begin{align*}
(\underbrace{x_{i_k}x_{i_{k-1}}\cdots x_{i_1}}_\eta)\circ d
&=\psi_d(x_{i_k}x_{i_{k-1}}\cdots x_{i_1})(\mbf{1}) \\
&=\psi_d(x_{i_k})\opercomp \psi_d(x_{i_{k-1}})\opercomp\cdots \opercomp \psi_d({x_{i_1}})(\mbf{1}).
\end{align*}
Linearly extend the product in its left argument into a product on $\DseriesXc{{\rm d}}\times\DseriesXd{{\rm d}}$ so that
\begin{align*}
c\circ d&=\sum_{\eta\in X_c^\ast} (c,\eta)\,\eta\circ d \\
&=\sum_{\eta\in X_c^\ast} (c,\eta)\,\psi_d(\eta)(\mbf{1}).
\end{align*}

\begex
Continuing Example~\ref{ex:simple-linear-composition}, it follows that
\begdi
c\circ d=(c,x_c)\psi_d(x_c)({\mbf 1})=(c,x_c)x_0(d\shuffle \mbf{1})=(c,x_c)x_0d=(c,x_c)(d,x_d)x_0x_d,
\enddi
so that
\begdi
(c\circ d)(\theta)=(c,x_c)(\theta)(d,x_d)(\theta)x_0x_d=\theta\frac{\partial}{\partial\theta}\theta^2\frac{\partial}{\partial\theta}x_0x_d
\enddi
as expected.
\endex

The following theorem summarizes the situation for the case above.

\begth \label{th:pd-CFS-series-generating-series}
Define the alphabets $X_c=\{x_0,x_c\}$ and $X_d=\{x_0,x_d\}$. Let
$c\in\DseriesXc{{\rm d}}$ and $d\in\DseriesXd{{\rm d}}$ be the generating series for
$y_c=F_c[u_c]$ and $y_d=F_d[u_d]$, respectively. If $F_c$ is linear, then
the series interconnection
\begdi
y(\theta,t)=(F_c\circ F_d) [u](\theta,t)
\enddi
has the representation $F_{c\circ d}$, where the composition product $c\circ d\in\DseriesXd{{\rm d}}$.
\endth

\begpr
It is first shown by induction on the length of the word
$\eta\in X_c^\ast$ that $E_{\eta}\circ F_d=F_{\eta\circ d}$
for any  $d\in\DseriesXd{{\rm d}}$.
Trivially,
\begdi
(E_{\emptyset}\circ F_d)[u]=E_{\emptyset}[F_d[u]]=E_{\emptyset}[u]=F_{\psi_d(\emptyset)(\mbfss{1})}[u].
\enddi
Next, assume that the claim holds for all words $\eta$ up to length $k\geq 0$.  Then
for any $x_i\in X_c$ observe that
\begin{align*}
E_{x_i\eta}[F_d[u]](t,t_0)&=\int_{t_0}^t F_{d_i}[u](\tau)\,
E_{\eta}[F_d[u]](\tau,t_0)\,d\tau \\
&=F_{x_0(d_i\shuffle \psi_d(\eta)(\mbfss{1}) )}[u](t) \\
&=F_{\psi_d(x_i\eta)(\mbfss{1})}[u](t),
\end{align*}
where $d_0=\mbf{1}$ and $d_c=d$.
Thus, the identity in question holds for every $\eta\in X_c^\ast$,
and all the shuffle products are well defined from the linearity assumption.
Finally, observe that
\begin{align*}
(F_c\circ F_d)[u]&=
\sum_{\eta\in X_c^{\ast}} (c,\eta)E_{\eta}[F_d[u]]
= \sum_{\eta\in X_c^{\ast}} (c,\eta) F_{\psi_d(\eta)(\mbfss{1})}[u] \\
&= \sum_{\eta\in X_c^{\ast}} (c,\eta) \left[\sum_{\nu\in X_d^{\ast}}
(\psi_d(\eta)(\mbf{1}),\nu)E_{\nu}[u] \right] \\
&= \sum_{\nu\in X_d^{\ast}}\left[\sum_{\eta\in X_c^{\ast}}(c,\eta) (\psi_d(\eta)(\mbf{1}),\nu)
\right] E_{\nu}[u] \\
&= \sum_{\nu\in X_d^{\ast}} (c\circ d,\nu)\: E_{\nu}[u] \\
&= F_{c\circ d}[u].
\end{align*}
\endpr

\begex
Integrating both sides of the transport equation \rref{eq:transport-equation} with respect to $t$ gives
\begeq \label{eq:integral-transport-eq}
\left(I+V\frac{\partial}{\partial\theta}E_{x_1}\right)[y]=y_0+E_{x_1}[u].
\endeq
Using the linearity of the operator
$I+V(\partial/\partial\theta)E_{x_1}$, its composition inverse can be written as
\begin{align*}
\left(I+V\frac{\partial}{\partial\theta}E_{x_1}\right)^{\circ -1}
&=\sum_{k=0}^\infty \left(-V\frac{\partial}{\partial\theta}E_{x_1}\right)^{\circ k} \\
&=\sum_{k=0}^\infty \left(-V\frac{\partial}{\partial\theta}\right)^{k}E_{x_1^{\circ k}} \\
&=I+\sum_{k=1}^\infty \left(-V\frac{\partial}{\partial\theta}\right)^{k}E_{x_0^{k-1}x_1}.
\end{align*}
Applying this inverse operator to both sides of \rref{eq:integral-transport-eq} yields
\begin{align*}
y&=\left(I+\sum_{k=1}^\infty \left(-V\frac{\partial}{\partial\theta}\right)^{k}E_{x_0^{k-1}x_1}\right)\circ(y_0+E_{x_1}[u]) \\
&=\sum_{k=0}^\infty \left(-V\frac{\partial}{\partial \theta}\right)^ky_0 E_{x_0^k}[1]+\sum_{k=0}^\infty \left(-V\frac{\partial}{\partial \theta}\right)^k E_{x_0^kx_1}[u],
\end{align*}
which is consistent with the solution given in \rref{eq:CF-solution-transport-eq}.
\endex

\begex
Reconsider the two single-parameter SISO transport systems in Example~\ref{ex:parallel-transport-systems} but now in
a series interconnection. The generating series for the composite system is
\begin{align*}
c\circ d&=\left(\sum_{k=0}^\infty \left(-V_c\frac{\partial}{\partial \theta_c}\right)^k x_0^kx_c\right)\circ
\left(\sum_{l=0}^\infty\left(-V_d\frac{\partial}{\partial \theta_d}\right)^l {x_0^lx_d}\right) \\
&=\sum_{k,l=0}^\infty \left(-V_c\frac{\partial}{\partial \theta_c}\right)^k
\left(-V_d\frac{\partial}{\partial \theta_d}\right)^l (x_0^kx_c\circ x_0^lx_d) \\
&=\sum_{k,l=0}^\infty \left(-V_c\frac{\partial}{\partial \theta_c}\right)^k
\left(-V_d\frac{\partial}{\partial \theta_d}\right)^l x_0^{k+l+1}x_d.
\end{align*}
In this case, linearity is preserved. Alternatively, starting with the system of equations
describing the interconnection,
\begdi
\left(\frac{\partial}{\partial t}+V_c\frac{\partial}{\partial \theta_c}\right)y=y_d,\;\;
\left(\frac{\partial}{\partial t}+V_d\frac{\partial}{\partial \theta_d}\right)y_d=u,
\enddi
it follows directly that
\begdi
\left(\frac{\partial}{\partial t}+V_d\frac{\partial}{\partial \theta_d}\right)
\left(\frac{\partial}{\partial t}+V_c\frac{\partial}{\partial \theta_c}\right)y=u.
\enddi
In light of the previous example,
\begin{align*}
y&=\left(\frac{\partial}{\partial t}+V_c\frac{\partial}{\partial \theta_c}\right)^{\circ -1}\circ
\left(\frac{\partial}{\partial t}+V_d\frac{\partial}{\partial \theta_d}\right)^{\circ -1}[u] \\
&=\left(I+\sum_{k=1}^\infty \left(-V_c\frac{\partial}{\partial\theta_c}\right)^{k}E_{x_0^{k-1}x_c}\right)\circ E_{x_c}\circ
\left(I+\sum_{l=1}^\infty \left(-V_d\frac{\partial}{\partial\theta_d}\right)^{l}E_{x_0^{l-1}x_d}\right)\circ E_{x_d}[u]\\
&=\left(\sum_{k=0}^\infty \left(-V_c\frac{\partial}{\partial\theta_c}\right)^{k}E_{x_0^lx_c}\right)\circ
\left(\sum_{l=0}^\infty \left(-V_d\frac{\partial}{\partial\theta_d}\right)^{l}E_{x_0^lx_d}[u]\right)\\
&=\sum_{k,l=0}^\infty \left(-V_c\frac{\partial}{\partial \theta_c}\right)^k
\left(-V_d\frac{\partial}{\partial \theta_d}\right)^l E_{x_0^{k+l+1}x_d}[u],
\end{align*}
as expected from the first analysis.
\endex

\begex
Consider the second-order PDE
\begeq \label{eq:2nd-order-PDE}
\frac{\partial^2 y}{\partial t^2}+\alpha_1\frac{\partial^2 y}{\partial t\partial \theta}+\alpha_2\frac{\partial^2 y}{\partial \theta^2}=u,
\endeq
where $(\theta,t)\in\re\times\re^+$ and $\alpha_i\in C^{\infty}(\re)$.
Assume the smooth initial conditions
$y(\theta,0)=y_0(\theta)$ and $(\partial{y}/\partial t)(\theta,0)=y_1(\theta)$.
Integrating \rref{eq:2nd-order-PDE} twice with respect to $t$ gives
\begdi
\left(I+\alpha_1 \frac{\partial}{\partial \theta}E_{x_1}+\alpha_2 \frac{\partial^2}{\partial \theta^2}E_{x_0x_1}\right)[y]=
y_0+\left(y_1+\alpha_1\frac{\partial}{\partial\theta}y_0\right)E_{x_0}[u]+E_{x_0x_1}[u].
\enddi
Therefore, \rref{eq:2nd-order-PDE} has the formal solution
\begeq \label{eq:sol-2nd-order-PDE}
y=\sum_{k=0}^\infty
\left(-\alpha_1 \frac{\partial}{\partial \theta}E_{x_1}-\alpha_2 \frac{\partial^2}{\partial \theta^2}E_{x_0x_1}\right)^{\circ k}\circ
\left(y_0+\left(y_1+\alpha_1\frac{\partial}{\partial\theta}y_0\right)E_{x_0}[u]+E_{x_0x_1}[u]\right).
\endeq
If the operator
\begdi
D:=I+\alpha_1\frac{\partial y}{\partial \theta}+\alpha_2\frac{\partial^2 y}{\partial \theta^2}
\enddi
is factorizable as a polynomial in ${\mathscr D}_1$ so that there exists $\beta_i\in{\mathscr D}_1$ (replacing the ground
field $\re$ with $\C$ if necessary)
such that
\begdi
D=\left(I+\beta_1\frac{\partial }{\partial\theta}\right) \left(I+\beta_2\frac{\partial }{\partial\theta}\right),
\enddi
then \rref{eq:sol-2nd-order-PDE}
can be rewritten as
\begdi
y=\sum_{k,l=0}^\infty
\left(-\beta_2\frac{\partial }{\partial\theta}\right)^k\left(-\beta_1\frac{\partial }{\partial\theta}\right)^l
\left[y_0E_{x_0^{k+l}}[u]+\left(y_1+\alpha_1\frac{\partial}{\partial\theta}y_0\right) E_{x_0^{k+l+1}}[u]
+E_{x_0^{k+l+1}x_1}[u]\right].
\enddi
That is, the system can be viewed as a series connection of two first-order systems.
Alternatively, when $\beta_1\neq \beta_2$, the partial fraction expansion
\begdi
D^{\circ -1}=\frac{\beta_1}{\beta_1-\beta_2}\left(I+\beta_1\frac{\partial }{\partial\theta}\right)^{\circ -1}+
\frac{\beta_2}{\beta_2-\beta_1}\left(I+\beta_2\frac{\partial }{\partial\theta}\right)^{\circ -1}
\enddi
gives the parallel sum representation
\begin{align*}
y&=\frac{\beta_1}{\beta_1-\beta_2}\sum_{k=0}^\infty
\left(-\beta_1\frac{\partial }{\partial\theta}\right)^k
\left[y_0E_{x_0^{k}}[u]+\left(y_1+\alpha_1\frac{\partial}{\partial\theta}y_0\right) E_{x_0^{k+1}}[u]
+E_{x_0^{k+1}x_1}[u]\right]\\
&\hspace*{0.2in}+\frac{\beta_2}{\beta_2-\beta_1}\sum_{k=0}^\infty
\left(-\beta_2\frac{\partial }{\partial\theta}\right)^k
\left[y_0E_{x_0^{k}}[u]+\left(y_1+\alpha_1\frac{\partial}{\partial\theta}y_0\right) E_{x_0^{k+1}}[u]
+E_{x_0^{k+1}x_1}[u]\right].
\end{align*}

Consider the specific case of a wave equation where $\alpha_1=0$, $\alpha_2=-1$, and $y_0=y_1=0$.
Then $\beta_1=1$ and $\beta_2=-1$ so that
\rref{eq:2nd-order-PDE} has the formal solution
\begdi
y=\sum_{k=0}^\infty
\frac{\partial^{2k}}{\partial \theta^{2k}}
E_{x_0^{2k+1}x_1}[u].
\enddi
In terms of a series connection, this is equivalent to
\begdi
y=\sum_{k,l=0}^\infty
(-1)^k\left(\frac{\partial }{\partial\theta}\right)^{k+l}
E_{x_0^{k+l+1}x_1}[u].
\enddi
As a parallel sum connection it has the representation
\begdi
y=\frac{1}{2}\sum_{k=0}^\infty
\left(-\frac{\partial }{\partial\theta}\right)^k
E_{x_0^{k+1}x_1}[u]
+\frac{1}{2}\sum_{k=0}^\infty
\left(\frac{\partial }{\partial\theta}\right)^k
E_{x_0^{k+1}x_1}[u].
\enddi
% zzz
%Suppose $u(\theta,t)=\exp(-t)\sin(\pi\theta)$. Then...\textcolor[rgb]{1.00,0.00,0.00}{check this against ICSTCC22 Example 4.3.}
\endex

%zzz
%\textcolor[rgb]{1.00,0.00,0.00}{Can we define relative degree in this setting?}

\section{Conclusions}

A class of parameter dependent Chen--Fliess series was introduced where the series coefficients are taken from a noncommutative ring of multivariable differential operators.
Sufficient conditions for convergence were given for a class of such series in the linear case.
It was also shown that these functional series are closed under the set of nonrecursive interconnections modulo some sufficient conditions for the series interconnection.
Specific examples are given involving the transport equation and the wave equation.

\vspace*{0.1in}

\section*{Acknowledgment}

The authors wish to thank Kurusch Ebrahimi-Fard and Fabian Harang for the
opportunity to participate in their workshops
supported by the project {\em Signatures for Images} at the Centre for Advanced Study at the
Norwegian Academy of Science and Letters in Oslo, Norway.
These events inspired much of the work presented here.

\end{document}